# Spin-orbit Interactions and the Nematicity observed in the Fe-based Superconductors

P.D. Johnson, H.-B. Yang, J. D. Rameau, G. D. Gu, Z.-H. Pan, and T. Valla

*Condensed Matter Physics and Materials Science Department,*

*Brookhaven National Laboratory, Upton, New York 11973, USA*

M. Weinert

*Department of Physics, University of Wisconsin-Milwaukee, Milwaukee, WI 53201, USA*

A. V. Fedorov

*Advanced Light Source, Lawrence Berkeley National Laboratory, Berkeley, CA 94720, USA*

(Dated: April 18, 2014)

Abstract

High resolution angle-resolved photoelectron spectroscopy (ARPES) is used to examine the electronic band structure of FeTe$_{0.5}$Se$_{0.5}$ near the Brillouin zone center. A consistent separation of the $\alpha_1$ and $\alpha_2$ bands is observed with little $k_z$ dependence of the $\alpha_1$ band. First-principles calculations for bulk and thin films demonstrate that the antiferromagnetic coupling between the Fe atoms and hybridization-induced spin-orbit effects lift the degeneracy of the Fe *dxz* and *dyz* orbitals at the zone center leading to orbital ordering. These experimental and computational results provide a natural microscopic basis for the nematicity observed in the Fe based superconductors.

The discovery of high $T_C$ superconductivity in the Fe-based materials [1,2] has led to a resurgence of interest in unconventional superconductivity in general. Like the superconducting cuprates discovered earlier, these new materials are essentially two-dimensional in character and have an underlying magnetic ground state in the parent compounds. The phase diagram in both systems suggests an interplay between magnetism and superconductivity. However, in the Fe-based materials the two ground states appear to co-exist over a small region of the phase diagram. One distinct difference between the two systems is that in the cuprates the superconductivity is achieved by doping a Mott insulator [3]; in the Fe-based the materials superconductivity emerges from a metallic normal state.

A much studied characteristic of the Fe-based superconductors is the tendency to show some form of electronic nematicity or orbital ordering [4]. This observation was first made spectroscopically in a scanning tunneling spectroscopy study (STS) of $Ca(Fe_{1−x}Co_x)2As2$ [5]. A combined neutron scattering/ARPES study [6] of the low-energy excitations in the FeTeSe system indicated a distinct polarization dependence in the spectral intensities of the electron pocket associated with the Fermi surface of this multiband system. Combined with the neutron scattering observations, the authors postulated some form of nematic order. In another ARPES study of the material, $Ba(Fe_{1-x}Co_x)2As_2$ [7,8], the authors reported the observation of the development of a symmetry breaking structural transition and subsequent magnetic ordering, a distinct polarization dependence, indicative of orbital ordering, was observed in the low-energy excitations. In a Point Contact Spectroscopic (PCS) study of the same system, Greene and co-workers identified a zero bias feature which they also associated with orbital ordering [9]. In a more recent STS study of $Ca(Fe_{1-x}Co_x)2As_2$ [10], the authors reported the observation of unidirectional electronic nanostructures as the system undergoes a symmetry-breaking structural transition at lower temperatures with the subsequent development of magnetic order. It is clearly important to understand these phenomena, which all develop in the normal state from which the superconductivity emerges. Several studies have previously proposed that the phenomena reflect some form of coupling between spin and orbital excitations.[6, 11] Here we examine this possibility directly by comparing ARPES experiments with first-principles calculations that include the spin-orbit interaction on all sites *including the anions*. In particular, we focus on studies of the Fe chalcogenide systems, $FeTe_{0.5}Se_{0.5}$ and $K(FeSe)_2$. We show that the spin-orbit interaction, which on the chalcogenides sites has an energy scale comparable to the exchange interaction between the Fe sites, can indeed be a pathway to the observed orbital ordering and nematicity.

For the experimental studies, single crystals of FeTe$_{0.5}$Se$_{0.5}$ were grown by a unidirectional solidification method. The nominal composition had no excess Fe, and T$c$, measured by magnetic susceptibility, is 14 K. Single crystals of KFe$_{2-y}$Se$_2$ were grown by the Bridgewater method and characterized by electron microscopy and X-ray diffraction. In these samples T$_c$ was measured to be 32K. The ARPES spectra were recorded on beamline U13UB at the National Synchrotron Light Source using a Scienta SES2002 electron spectrometer, and on beamline 12.0.1 at the Advanced Light Source using a Scienta SES100 electron spectrometer. The photoelectron energy and angular resolution were 7 to 15 meV and 0.1°, respectively. Samples were cleaved *in situ* (base pressure of $5 \times 10^{-11}$ Torr and held at T = 25 K for the measurements. STM studies of cleaved FeSe$_{0.5}$Te$_{0.5}$ indicate that the surface stoichiometry is close to that of the bulk.[12]

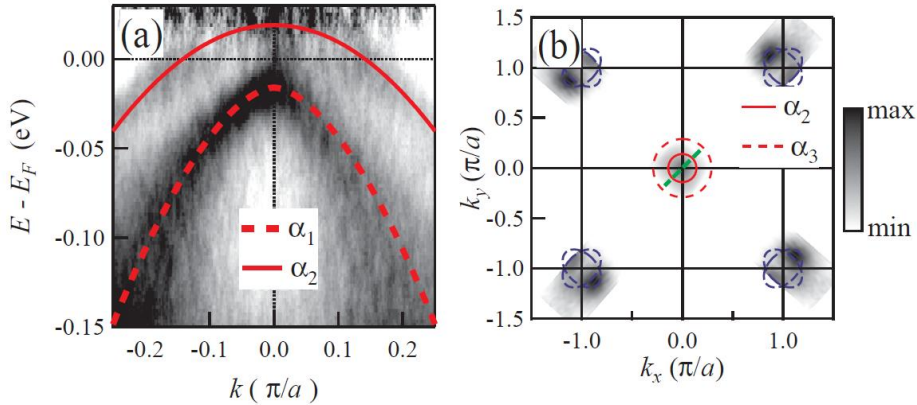

FIG. 1. (color online) ARPES spectrum of the *dxz* and *dyz* bands of FeTe0.5Se0.5. (a) Fermi function renormalized ARPES spectrum near the Brillouin zone center.(Γ point) The dispersion of the *α*1 band (dashed red) and *α*2 band (red) are traced by colored solid lines. (b) The ARPES spectral intensity at Fermi energy in the first Brillouin zone. The green line shows the location of the cut in (a), and the red circle traces the Fermi surface formed by the *α*2 band, the dashed red line indicates the Fermi surface formed by the α$_3$ band.

Figure 1 shows the experimentally measured bands near the center of the zone and the measured Fermi surface for the FeTe$_{0.5}$Se$_{0.5}$ system. The Fermi surface shows a characteristic hole-pocket at the center of the zone and electron pockets at the zone corners. Similar results have been reported elsewhere [6].6 In an early study Nakayama *et al.* also reported three bands near the Γ point, but with only two bands actually forming a Fermi surface [13]. This is in contrast to LDA calculations that routinely predict three Fermi surfaces in the vicinity of the Γ point [14]. In a subsequent photon energy dependent study of the same Fermi surface this discrepancy was ascribed to a dependence on $k_z$ [15]. However in the present study, high-resolution measurements of the Fermi surface as a function of incident photon energy clearly show that the $\alpha_2$ derived Fermi surface at the center of the zone has a radius of 0.15 $\text{Å}^{-1}$, independent of $k_z$. $\alpha_1$ band is always below the Fermi level, with the top of the band always at about 20 meV below the Fermi level. As we discuss later, these two bands, $\alpha1$ and $\alpha2$, are the results of hybridization of the Fe $d_{xz}$ and $d_{yz}$ states.

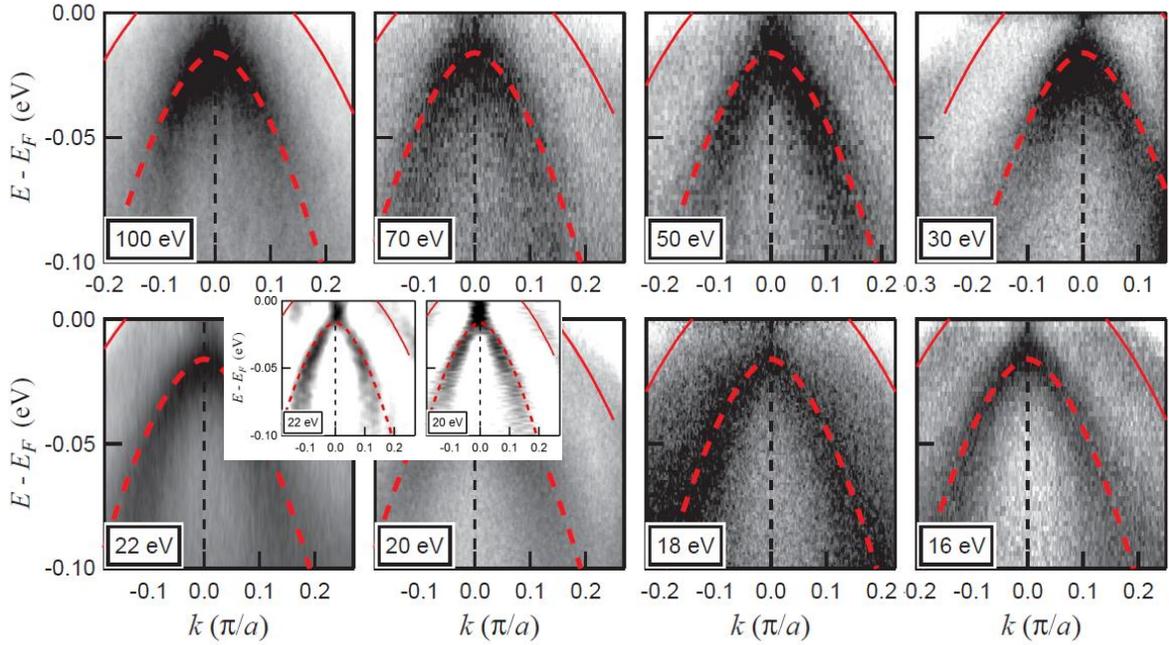

FIG. 2. (color online) Photon energy dependence of intensity of the $\alpha1$ and $\alpha2$ bands measured near the Γ point. (Spectra are renormalized by the Fermi function.) The inset shows second derivative plots for the incident photon energies of 22 and 20 eV. The measurements at different photon energies correspond to $k_z$ points in units of $2\pi/c$ as indicated in the brackets: 16 eV [0.57], 18 eV [0.71], 20 eV [0.83], 22 eV [0.95], 30 eV [0.36], 50 eV [0.17][, 70 eV, [0.81], 100 eV [0.61].

Figure 2 shows representative results from a photon energy dependent study. Although the relative intensity of the $\alpha_1$ band differs due to matrix element, including polarization, effects, little $k_z$ dispersion is observed when the photon energy is varied. Furthermore, there is no evidence of the existence of a Fermi surface derived from the $\alpha_1$ band. To highlight the dispersion of the $\alpha_2$ band we show representative derivative plots in the inset. As indicated in figure 1, there is a further band, $\alpha_3$, that crosses the Fermi level and forms a hole Fermi surface around the Γ point with a radius of 0.3 Å$^{-1}$. The $\alpha_3$ band is mainly derived from the $d_{x^2-y^2}$ orbital. Unlike the $\alpha_1$ and $\alpha_2$ bands, our observations for the $\alpha_3$ band are consistent with previous reports [16-18]. The present study shows that the $\alpha_1$ and $\alpha_2$ bands are highly two-dimensional, in contrast to previous ARPES results [15]. Furthermore, our ARPES results show that the $\alpha 1$ and $\alpha 2$ bands have very similar band widths and a similar intensity dependence with photon energy, which leads us to conclude that they reflect hybridization of the $d_{xz}$ and $d_{yz}$ orbitals, which, in the absence of any other symmetry breaking, are strongly mixed. In contrast, although not shown here, the $\alpha 3$ band has a much narrower band width and a different matrix element sensitivity, consistent with the characteristics of the $d_{x2-y2}$ orbital. The observations are similar to the earlier work by Tamai [16]. However in contrast to the latter study, we assign different orbital character to the bands near the Γ point.

In the following we re-examine these experimental observations in the light of calculations that include the spin-orbit interaction on all sites. The calculations were carried out using the Full-potential Linearized Augmented Plane Wave (FLAPW) method as implemented in *flair* [19]. Structural models of both bulk and surfaces corresponding to various configurations and concentrations of Te and Se were considered, with most of the calculations carried on for FeTe$_{0.5}$Se$_{0.5}$. The bulk volumes and positions were relaxed. The wave functions and density/potential cutoffs were 220 and 2000 eV, respectively; the PBE form of the GGA was used for exchange-correlation. The Brillouin zone was sampled with k-point meshes equivalent to 4800 point for the primitive 4-atom unit cell. Bulk defect calculations used supercells of up to $4 \times 4 \times 2$, while surface calculations used up to $2 \times 2$ supercells in-plane, and 5–9 Fe(Te,Se) (tri-)layers.

Different magnetic configurations were calculated starting from initial spin arrangements and then relaxed. In addition, fixed moment calculations were performed, both to ensure that local solutions were not missed and to enforce zero net moments for models of the paramagnetic state. The calculations strongly support the conclusion that there will be local moments on the Fe atoms (~2 $\mu_B$), and that these moments in bulk Fe$_2$(Te,Se) prefer to be antiferromagnetically (AFM) coupled; in fact, even in fixed moment calculations, no stable (zero-field) ferromagnetic state was found. On the other hand, various stable AFM configurations were found, with the observed stripe ordering (whose magnetic cell is twice the chemical cell) slightly more stable than the simple AFM ordering of the two Fe atoms in the primitive unit cell (where the magnetic and chemical primitive cells are the same). Local spin flips, such as expected in the paramagnetic state at higher temperatures, introduce additional FM couplings between Fe atoms, but these do not destroy the magnetic state. Although the particular magnetic configurations, the distribution of Te and Se atoms, defects (vacancies, anti-sites, etc.), and the presence of surfaces do affect details of the calculated electronic states, we will focus on the (robust) commonalities that relate to the experimental data.

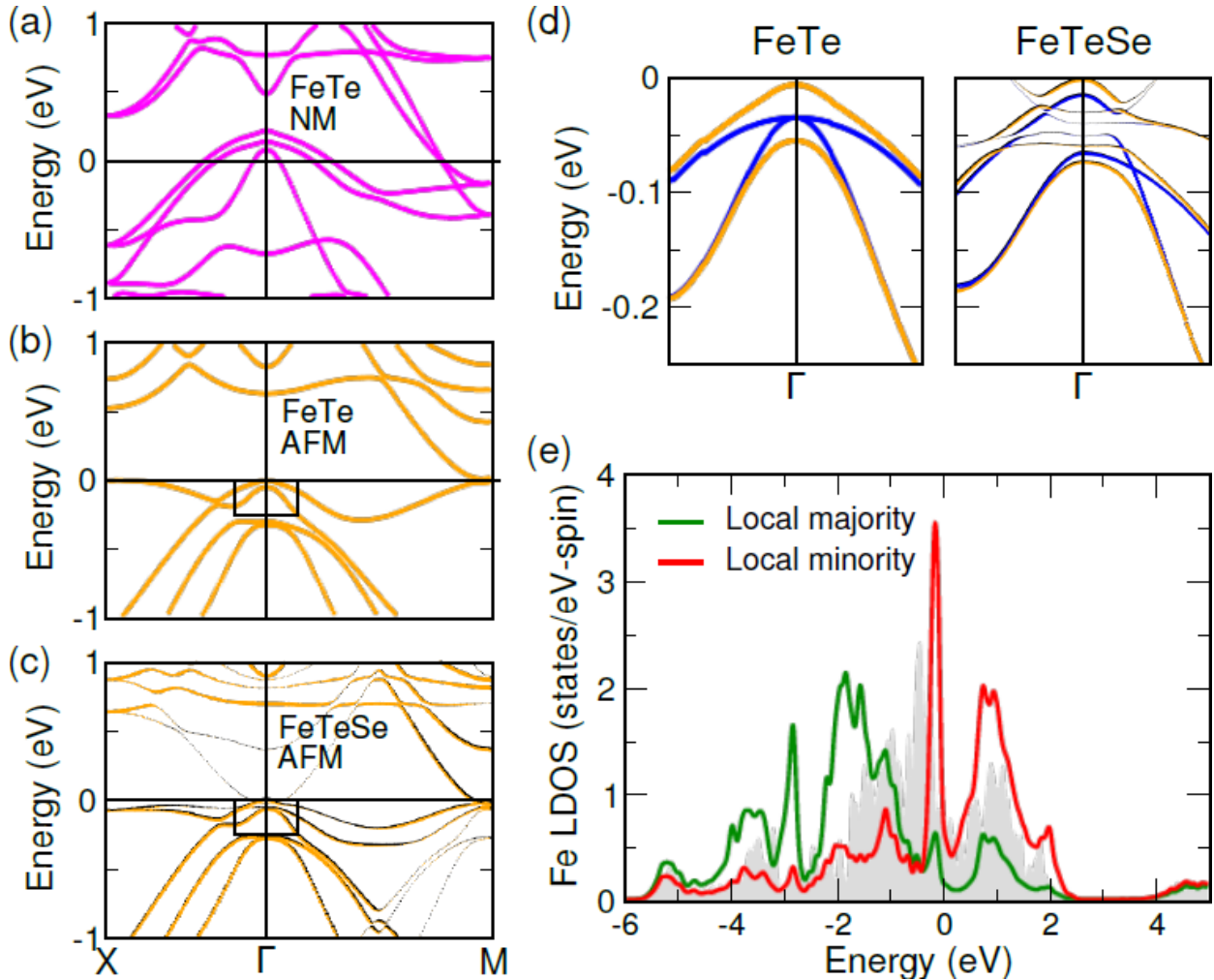

FIG. 3. (color online) Spin-orbit bands along the high-symmetry lines of the (1×1) Brillouin zone for (a) non-magnetic (NM) FeTe, (b) antiferromagnetic (AFM) FeTe, and (c) AFM FeTe$_{0.5}$Se$_{0.5}$ with the Te and Se in a c(2×2) ordering. (d) Expanded views of the bands around Γ in boxed regions in (b) and (c); bands with (without) spin-orbit in orange (blue), and the the size of the dots gives the k-projection weight. (e) Spin-resolved local Fe density of states for (c); the gray background is the corresponding NM LDOS.

Although the systems are magnetic, even the bulk non-magnetic (NM) FeTe bands (Fig. 3a) are strongly modified around $\Gamma$ by spin-orbit effects, including significant splittings of the states. (Because Te is heavier than Se, the spin-orbit effects are more noticeable in FeTe than in FeSe.) However, the relative position of the NM bands to the Fermi level are clearly not compatible with the experimental results, further indicating the importance of magnetic interactions in these systems. The combined effects of spin-orbit, magnetic ordering, and structure/Te-Se configuration can be seen in bulk bands of FeTe and FeTeSe, Figs.3b,c. Allowing AFM coupling causes a shift downward of the states around $\Gamma$ and upward around M for both FeTe and FeTeSe. Despite Te and Se having the same valence, the dispersion and relative placement of the bands around the zone boundaries are modified by the Te/Se substitution; in particular, the inclusion of Se leads to a lowering of the valence band edges along $\Gamma$-X, and also smaller shifts around M, reflecting the difference in spatial extent of the Te and Se orbitals which leads to changes in the hybridization with the Fe.

For FeTe around $\Gamma$ (Fig. 3d), the bands without spin-orbit for the two spin directions are degenerate because of the 4-fold magnetic space group symmetry relating the two Fe atoms. With spin-orbit included, the states at $\Gamma$ are split by ~60 meV. This effect, including the degeneracies, is also expected for FeTe$_{1-x}$Se$_x$ if the Se substitute randomly on the Te site *and* the system can be described well by the averaged system, i.e., short-range correlations/orderings are not dominant. Replacing half the Te by Se in each layer introduces a symmetry breaking that already splits the degeneracy at $\Gamma$. With spin-orbit, the splitting at $\Gamma$ is about the same size, but there are now three bands along $\Gamma$-M, with the uppermost flat band rather flat. Comparing the two situations in Fig. 3d, as well as other bulk calculations [6,11], it becomes clear that spin-orbit plays an important role in determining the topology of the bands around $\Gamma$, especially splittings, with the observation that the spin-orbit bands for different configurations are more similar than are the bands that neglect it. This splitting of the bands at $\Gamma$ with spin-orbit, and the breaking of the degeneracy between the $d_{xz,yz}$ orbitals, can be understood in part from simple symmetry arguments: the site symmetry of the Fe atoms is (at most) D$_{2d}$, which has a 2-fold irreducible representation that splits under spin-orbit (double group).

One important consequence of the short-range AFM correlations is that the interaction between Fe orbitals of a given type (e.g., $d_{xz,yz}$) on neighboring AFM coupled Fe sites are between local majority/minority states: because of the local exchange splitting, these orbitals differ, i.e., the $d_{xz,yz}\uparrow$ orbitals on one site interacting with the $d_{xz,yz}\downarrow$ orbitals on the other are not the same. This spin-dependent $_{difference}$ is also seen in the Fe local density of states (LDOS) projected onto different spins, Fig. 3e. The majority and minority channels clearly differ significantly, demonstrating that models of these materials should not assume rigidly shifted local bands. A common feature in the LDOS for different spin configurations is the sharp peak in the minority spin channel in the vicinity of the Fermi level, and a corresponding – but weaker – peak for majority spin. From the calculated (local) density matrix, the minority spin has a large $d_{z^2}$ contribution, as well as $d_{x^2-y^2}$ and $d_{xz,yz}$, while the minority peak is mainly $d_{xz,yz}$, consistent with the experimental observations. More importantly, however, is the fact that the $d_{xz}$ and $d_{yz}$ contributions to the density matrix around the Fermi level show a strong asymmetry of around 17% for this calculation, even though the calculated system has a C4 magnetic operation. This behavior is not peculiar to this particular calculation, but is a common feature of our other calculations. Thus, the orbital ordering/nematicity appears as a natural consequence of magnetic and spin-orbit

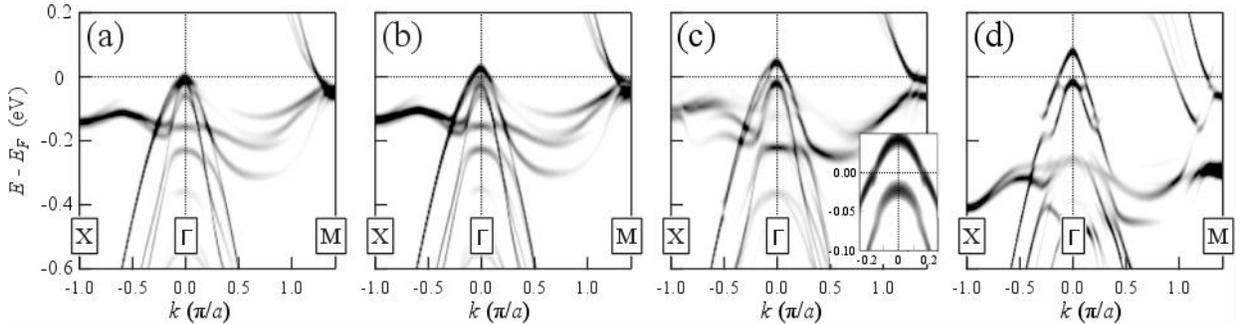

interactions in these systems.

FIG. 4. Calculated k-projected surface bands for a 9 Fe-Te-Se layer film in a c(2×2) configuration with equal number of Te and Se in each plane (a) without SOC and (b) with spin-orbit included. (c) Spin-orbit bands for p(2×2) 8-layer film with the top trilayer having a 5:3::Te:Se ratio; the insert is an expanded view around Γ along the same direction as the experiments in Fig. 2. (d) Same as in (b), but with the top trilayer having only Te.

To make more direct connection between the experiments and the calculations, we have modeled the surface using a finite slab geometry, and various Se-Te distributions to calculate the (supercell) bands. The plotted bands are shown relative to the first Brillouin zone of the primitive Fe(Te,Se) cell using a *k*-projection technique (and wave function weight in the surface region) to mimic the photoemission results and to allow coparisons among the different calculations and experiments. (This technique does an exact unfolding for pure supercells.) To account for the possibility of different orientations and domains, the bands along the different high-symmetry lines were calculated and then overlaid.

Figure 4 shows a calculation of the electronic structure of the FeTeSe system with and without the spin-orbit interaction included. As in the bulk, spin-orbit causes splittings and lifting of the degeneracy at the center of the zone approximately independent of the Se-Te configuration, including for the full Te case (Fig. 4d) where the non-spin-orbit bands at degenerate at $\Gamma$. The presence of the surface, spin-orbit, and the Te-Se concentration affect the position of the bands at $\Gamma$ and M relative to the Fermi level and also the localization to the surface trilayer. The overall comparison of the ARPES (Fig. 2) and the calculated bands (cf., inset to Fig. 4c) is reasonable. As in the case of bulk, the zone-center splitting reflects the hybridization of the Fe $d_{xz}$, $d_{yz}$ orbitals with the *p*-orbitals on the neighboring chalcogenide atoms. The spin-orbit splitting of the latter *p*-orbitals is large enough to lift the degeneracy observed at the center of the zone. (The Fe spin-orbit strength is smaller than the chalcogenide.) Note a similar hybridization-induced splitting is not expected to be easily observable in the cuprate system since the spin-orbit due to the oxygens is orders of magnitude smaller.

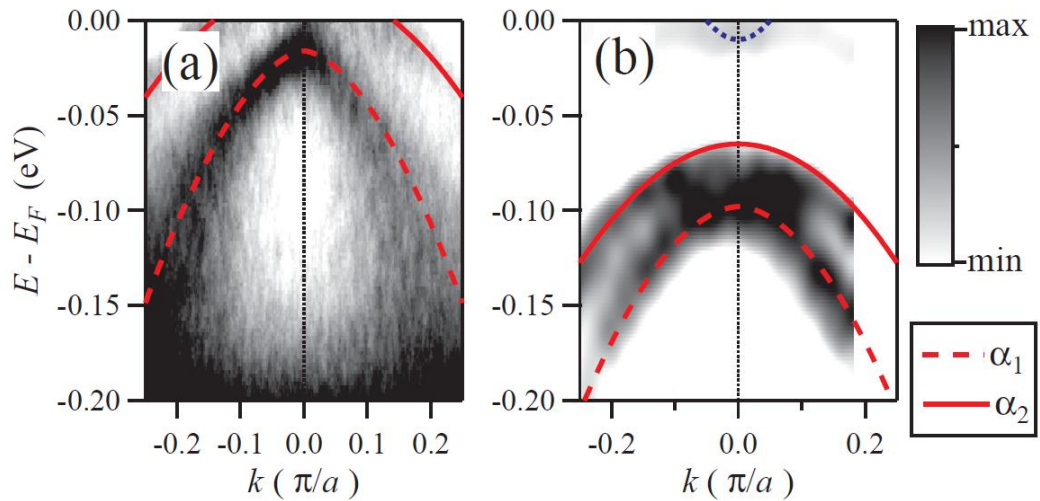

FIG. 5. (color online) Effect of the electron doping by K. Comparison of the spectra of FeTe$_{0.5}$Se$_{0.5}$ (a) and K(FeSe)$_2$ (b). Red line traces the dispersion of the $\alpha_2$ band, and dashed red line traces the dispersion the $\alpha_1$ band.

As in the bulk, there is an inherent asymmetry between the $d_{xz}$ and $d_{yz}$ orbitals, which will be enhanced by symmetry-breaking transitions, resulting in the observed "orbital ordering" phenomena. The calculated spin-orbit induced splitting of approximately 50 meV at the center of the zone, shown in Fig. 4c, compares favorably with the measured splitting of approximately 40 meV shown in Fig. 1a. (As seen in Fig. 4, lower Te concentrations in the surface result in a smaller spin-orbit–induced splitting, but will also result in small shifts of the states relative to the Fermi level. Likewise, defects, possible spin-flips at finite temperatures, etc., will modify the details of the calculated bands, but the qualitative features due to spin-orbit will also be present in those cases.

Figure 5 shows a comparison of the zone center photoemission spectra recorded from the same FeTe$_{0.5}$Se$_{0.5}$ system and the related K(FeSe)$_2$ system. The latter may be viewed as a similar Fe-based superconductor, but now with K intercalated between the superconducting Fe-chalcogenide layers. To be discussed in more detail elsewhere, we show this comparison because it clearly reveals the spin-orbit–induced splitting at the center of the zone as the charge donated by the alkali atom results in an approximately rigid band shift; the splitting is smaller in for K(FeSe)$_2$ than for FeTe$_{0.5}$Se$_{0.5}$ discussed here because the spin-orbit parameter of Se is smaller than Te.

In summary our experiments and calculations show that the spin-orbit interaction provides a natural explanation of the nematicity observed in the Fe-based superconductors, although the degree of "orbital ordering" is not complete. The relatively large spin-orbit interaction on the chalcogenide atoms (especially Te) in these systems lifts the degeneracy of the $d_{xz}$, $d_{yz}$ orbitals at the center of the zone. The spin-orbit interaction in the Fe $d$-orbitals alone is not enough to lift the degeneracy to the degree observed. While the hybridization lifts the degeneracy the bands are still mixed and to produce the observed "orbital ordering" [8,20] requires some other symmetry breaking phenomena, such as short-range chemical ordering during growth or magnetic ordering. However with orbital ordering in place, we can expect profound effects in t r a n s p o r t , tunneling and spectroscopy measurements.

Work at Brookhaven is supported by the Center for Emergent Superconductivity, an Energy Frontier Research Center funded by the U.S. DOE, Office of Basic Energy Sciences, and at UWM by the National Science Foundation (DMR-1335215).